\documentclass[11pt,graphicx,amsmath]{article}
\usepackage{amsmath}
\usepackage{graphicx}
\usepackage{bm}

\usepackage{amssymb}

\title{ Energy Momentum Pseudo-Tensor of Relic Gravitational Wave
                  in Expanding Universe }

\author{\small Daiqin Su and   Yang  Zhang \thanks{yzh@ustc.edu.cn}\
                        \\
    \small Key Laboratory for Researches in Galaxies and Cosmology, \\
     \small Department of  Astronomy,  University of Science and Technology of China, \\
     \small Hefei, Anhui, 230026,  China     }
 \date{}

\begin{document}

\maketitle
\baselineskip=19truept

\def\vek{\vec{k}}
\renewcommand{\arraystretch}{1.5}
\newcommand{\be}{\begin{equation}}
\newcommand{\ee}{\end{equation}}
\newcommand{\ba}{\begin{eqnarray}}
\newcommand{\ea}{\end{eqnarray}}

\sf

\begin{center}
\Large  Abstract
\end{center}

\begin{quote}

 \sf
\baselineskip=19truept

We study
the energy-momentum pseudo-tensor of gravitational wave,
and examine
the one introduced by Landau-Lifshitz for a general gravitational field
and the effective one recently used in literature.
In short wavelength limit after Brill-Hartle average,
both lead to the same gauge invariant stress tensor of gravitational wave.
For relic gravitational waves
in the expanding universe,
we  examine two forms of pressure, $p_{gw}$ and $\mathcal{P}_{gw}$,
and trace the origin of their difference
to a coupling between gravitational waves
and the background matter.
The difference is shown to be negligibly small
for most of cosmic expansion stages starting from inflation.
We demonstrate that the wave equation is equivalent
to the energy conservation equation using
the pressure  $\mathcal{P}_{gw}$
that includes the mentioned coupling.

\end{quote}


\begin{center}
{\bf 1. Introduction}
\end{center}

Relic gravitational waves (RGW) is believed to be
generated during the inflationary stage of the expanding universe,
and distributed as a stochastic background \cite{grishchuk,starobinsky,abbott,allen}.
Its amplitude, spectral index, running index, etc,
are mainly determined by the initial condition,
carrying a wealth of physical information of
very early universe \cite{Zhang}.
As a possible channel for its detection,
RGW can induce the curl type of polarization of
cosmic microwave background radiation (CMB) \cite{Basko,Seljak,Kamionkowski,ZhangCMB}.
Since RGW is important for the early universe,
one natural question is what are the energy density and pressure of RGW.
Different definitions of
the energy-momentum tensor of RGW lead to different equation of state
for long wavelength modes,
and thereby yield different predictions on
its cosmological influences \cite{Ford,Giovannini2}.
The field equation of RGW has been commonly employed as a solid ground
to study the evolution of RGW in the expanding universe,
and we  shall examine which definition corresponds
to the wave equation of RGW,
providing another perspective to the issue of energy and pressure of RGW.
In most cases for physical systems,
field equation and energy conservation equation are
two different, yet equivalent, ways to describe the systems.
As is known,
the field equation can be derived
from the equation of energy conservation, and vice versa,
for such systems as oscillator, electromagnetic wave in vacuum,
scalar field (both in flat and curved background spacetimes).
The energy and pressure of these systems are well defined,
and one can work with either equation for practical conveniences.
For gravitational field,
the problem of the energy-momentum is a thorny issue.
As it stands, the energy-momentum of gravitational field
can only be defined
as a pseudotensor $t^{\alpha\beta}$, in stead of a tensor.
This is due to its non-locality due to Einstein's equivalence principle  \cite{MTW}.
The disadvantage of  pseudotensor $t^{\alpha\beta}$  is that
it is not unique.
Various forms  have been proposed for $t^{\alpha\beta}$,
each being based on certain specific considerations,
such as the conservation of total energy momentum,
symmetry, additivity, and so on
\cite{MTW,Landau,Weinberg,Einstein,Moller1,Moller2,Bergmann,Papapetrou,Isaacson1}.

In this paper,
we shall study two definitions of
the energy momentum pseudo-tensor (EMPT) of gravitational waves,
emphasizing their relation to the wave equation.
The paper is organized as follow.
In section 2,
we first examine Landau-Lifshitz' definition $t^{\alpha\beta}$ of
EMPT of a general gravitational field,
and,  for the simple case of for the flat spacetime,
we demonstrate that the energy conservation of this $t^{\alpha\beta}$
leads to the field equation of gravitational waves (GW), and vice versa.
In section 3, we investigate
the short wavelength limit of Landau-Lifshitz'  $t^{\alpha\beta}$,
and apply Brill-Hartle average to it.
In section 4, we examine the effective EMPT $t^{\alpha\beta}_{eff}$ of GW
that appears in literature \cite{Isaacson1,Mukhanov1,Mukhanov2,Giovannini2}.
In the short wavelength limit after a spatial average,
$t^{\alpha\beta}_{eff}$ leads to the same  expression
as Landau-Lifshitz' does.
In section 5,
we study two definitions of the pressure, $p_{gw}$ and $\mathcal{P}_{gw}$,
of RGW in the expanding universe,
trace the origin of their difference
and reveal its physical meaning and cosmological implications.
Finally, in section 6,
for RGW  in the expanding Robertson-Walker (RW) spacetime,
we examine the relation between the wave equation
and the equation of energy conservation for the effective EMPT.

\begin{center}
{\bf 2. Landau-Lifshitz'   $t^{\alpha\beta}$ and Wave Equation}
\end{center}

For Landau-Lifshitz' definition of EMPT,
we will explicitly show that the energy conservation
is equivalent the field equation of GW
in flat spacetime.
For a general curved spacetime with the metric $g_{\alpha\beta}$,
Landau-lifshitz' definition of the energy-momentum pseudo-tensor
$t^{\alpha\beta}$ of gravitational field is given by \cite{Landau}
\ba \label{Landau-Lifshitz}
16 \pi G\, (-g) t^{\alpha\beta} &=&{\bf g}^{\alpha\beta}\,_{,\lambda}{\bf g}^{\lambda\mu}\,_{,\mu}
-{\bf g}^{\alpha\lambda}\,_{,\lambda}{\bf g}^{\beta\mu}\,_{,\mu}
+\frac{1}{2}g^{\alpha\beta}g_{\lambda\mu}{\bf g}^{\lambda\nu}\,_{,\rho}{\bf g}^{\rho\mu}\,_{,\nu}\nonumber \\
&&-(g^{\alpha\lambda}g_{\mu\nu}{\bf g}^{\beta\nu}\,_{,\rho}{\bf g}^{\rho\mu}\,_{,\lambda}
+ g^{\beta\lambda}g_{\mu\nu}{\bf g}^{\alpha\nu}\,_{,\rho}{\bf g}^{\rho\mu}\,_{,\lambda})
+g^{\nu\rho}g_{\lambda\mu}{\bf g}^{\alpha\lambda}\,_{,\nu}{\bf g}^{\beta\mu}\,_{,\rho}
\nonumber \\
&&+\frac{1}{8}(2g^{\alpha\lambda}g^{\beta\mu}-g^{\alpha\beta}g^{\lambda\mu})
(2g_{\nu\rho}g_{\sigma\tau}-g_{\rho\sigma}g_{\nu\tau})
{\bf g}^{\nu\tau}\,_{,\lambda}{\bf g}^{\rho\sigma}\,_{,\mu}
\ea
where ${\bf g}^{\alpha\beta}=(-g)^{1/2}g^{\alpha\beta}$,
$g=det\parallel g_{\alpha\beta}\parallel=det\parallel
{\bf g}^{\alpha\beta}\parallel$,
and the sub comma ``," stands for the ordinary derivative.
Einstein equation takes the form
\be \label{H}
H^{\alpha\mu\beta\nu}\, _{,\mu\nu}=16\pi(-g)(T^{\alpha\beta}+t^{\alpha\beta}),
\ee
where  $H^{\alpha\mu\beta\nu}
={\bf g}^{\alpha\beta}{\bf g}^{\mu\nu}-{\bf g}^{\alpha\nu}{\bf g}^{\beta\mu}$,
antisymmetric in $\beta$ and $\nu$,
and  $T^{\alpha\beta}$ is the energy-momentum tensor of the matter.
From Eq.(\ref{H}) it follows that
\be  \label{conservation}
[(-g)(T^{\alpha\beta}+t^{\alpha\beta})]_{,\beta}=0
\ee
is identically satisfied.
It implies that the four-momentum of the system,
i.e, the gravitational field plus the matter,
is conserved.
This equation implicitly allows for interactions between the two components.

For a general spacetime,
the expression  $ t^{\alpha\beta}$
in Eq.(\ref{Landau-Lifshitz})  is rather complicated.
We consider the flat spacetime background with
the metric $\eta_{\mu\nu}=diag(-1,1,1,1)$
perturbed by GW, defined by
$\bar{h}^{\alpha\beta}\equiv -{\bf g}^{\alpha\beta}+\eta^{\alpha\beta}$.
Then the pseudo-tensor reduces to
\ba \label{tper}
16\pi G\, (-g) t^{\alpha\beta}&=&\bar{h}^{\alpha\beta}\,_{,\lambda} \bar{h}^{\lambda\mu}\,_{,\mu}
-\bar{h}^{\alpha\lambda}\,_{,\lambda}\bar{h}^{\beta\mu}\,_{,\mu}
+\frac{1}{2}\eta^{\alpha\beta}\eta_{\lambda\mu}
          \bar{h}^{\lambda\nu}\,_{,\rho}\bar{h}^{\rho\mu}\,_{,\nu}\nonumber\\
&&-(\eta^{\alpha\lambda}\eta_{\mu\nu} \bar{h}^{\beta\nu}\,_{,\rho} \bar{h}^{\rho\mu}\,_{,\lambda}
   +\eta^{\beta\lambda}\eta_{\mu\nu} \bar{h}^{\alpha\nu}\,_{,\rho} \bar{h}^{\rho\mu}\,_{,\lambda})\nonumber\\
&&+\frac{1}{8}(2\eta^{\alpha\lambda}\eta^{\beta\mu}-\eta^{\alpha\beta}\eta^{\lambda\mu})
(2\eta_{\nu\rho}\eta_{\sigma\tau}-\eta_{\rho\sigma}\eta_{\nu\tau})
        \bar{h}^{\nu\tau}\,_{,\lambda} \bar{h}^{\rho\sigma}\,_{,\mu}
\ea

As a physical quantity,
GW has two degrees of freedom (polarizations),
while the perturbation $\bar{h}^{\alpha\beta}$ has ten components,
containing eight degrees of non-physical, gauge freedom.
To remove these from $\bar{h}^{\alpha\beta}$,
we choose a synchronous coordinate system, in which
\be \label{synchr}
\bar{h}^{0\beta}=0,
\ee
and, furthermore,
impose upon $\bar{h}^{ij}$ the traceless-transverse (TT) condition
as the following:
\be \label{TT}
\bar{h}^i\, _i=0, \,\, \bar{h}^{ij}\, _{,j}=0.
\ee
Eq.(\ref{synchr}) together with Eq.(\ref{TT})
are eight constraints.
Then  Eq.(\ref{tper}) reduces to
\ba \label{tab}
16\pi G\, t^{\alpha\beta}
&=&\frac{1}{2}\eta^{\alpha\beta}\bar{h}^{ij}\,_{,k}\bar{h}^{k}\,_{i,j}
+\bar{h}^{\alpha}\,_{i,\nu}\bar{h}^{\beta i,\nu}
-\bar{h}^{\alpha}\,_{i,j}\bar{h}^{ij,\beta}
-\bar{h}^{\beta}\,_{i,j}\bar{h}^{ij,\alpha}\nonumber\\
&&+\frac{1}{2}\bar{h}^{ij,\alpha}\bar{h}_{ij}\,^{,\beta}
-\frac{1}{4}\eta^{\alpha\beta}\bar{h}^{ij,\lambda}\bar{h}_{ij,\lambda}.
\ea
In absence of matter $T^{\alpha\beta}=0$,
Eq.(\ref{conservation}) becomes
 (keeping up to the second order of ${\bar h}^{\alpha\beta}$)
\be
t^{\alpha\beta}\, _{,\beta}=0,
\ee
i.e, the four-momentum of GW is conserved.
We will check that this equation is consistent with
the wave equation of GW.
We take the derivative of $t^{\alpha\beta}$ in Eq.(\ref{tab}):
\begin{eqnarray}
16\pi G \, t^{\alpha\beta}\, _{,\beta}
&=&\frac{1}{2}\bar{h}^{j,\alpha}_{i,k} \bar{h}^{k i}_{,j}+\frac{1}{2} \bar{h}^{j}_{i,k} \bar{h}^{ki,\alpha}_{,j}+\bar{h}^{\alpha}_{i,\nu j} \bar{h}^{ij,\nu}-\bar{h}^{\beta}_{i,j} \bar{h}^{ij,\alpha}_{,\beta}\nonumber\\
&&-\bar{h}^{\alpha}_{i,j\beta} \bar{h}^{ij,\beta}-\bar{h}^{\alpha}_{i,j} \bar{h}^{ij,\beta}_{,\beta}+\frac{1}{2} \bar{h}^{ij,\alpha}_{,\beta} \bar{h}^{,\beta}_{ij}
+\frac{1}{2} \bar{h}^{ij,\alpha} \bar{h}^{,\beta}_{ij,\beta}\nonumber\\
&&-\frac{1}{4} \bar{h}^{ij,\lambda \alpha} \bar{h}_{ij,\lambda}
-\frac{1}{4} \bar{h}^{ij,\lambda} \bar{h}^{,\alpha}_{ij,\lambda}.
\end{eqnarray}
For the $``\alpha=0"$ component, it reads as
\begin{eqnarray} \label{con}
16\pi G\, t^{0\beta}\, _{,\beta}
&=&\frac{1}{2}\bar{h}^{j,0}_{i,k} \bar{h}^{k i}_{,j}+\frac{1}{2} \bar{h}^{j}_{i,k} \bar{h}^{ki,0}_{,j}+\bar{h}^{0}_{i,\nu j} \bar{h}^{ij,\nu}-\bar{h}^{\beta}_{i,j} \bar{h}^{ij,0}_{,\beta}\nonumber\\
&&-\bar{h}^{0}_{i,j\beta} \bar{h}^{ij,\beta}-\bar{h}^{0}_{i,j} \bar{h}^{ij,\beta}_{,\beta}+\frac{1}{2} \bar{h}^{ij,0}_{,\beta} \bar{h}^{,\beta}_{ij}
+\frac{1}{2} \bar{h}^{ij,0} \bar{h}^{,\beta}_{ij,\beta}\nonumber\\
&&-\frac{1}{4} \bar{h}^{ij,\lambda 0} \bar{h}_{ij,\lambda}
-\frac{1}{4} \bar{h}^{ij,\lambda} \bar{h}^{,0}_{ij,\lambda}\nonumber\\
&=&\frac{1}{2} \bar{h}^{ij,0}\bar{h}_{ij}\, ^{,\mu}\, _{,\mu}.
\end{eqnarray}
From this equality it follows that
the energy conservation
\be
t^{0\beta}\,  _{,\beta}=0
\ee
implies the wave equation of GW
\be
\bar{h}_{ij}\, ^{,\mu}\, _{,\mu}=0,
\ee
and vice versa.

\begin{center}
{\bf 3. Short Wavelength Limit of Landau and Lifshitz' $t^{\alpha\beta}$}
\end{center}

We are interested in GW in curved spacetimes,  such as in an expanding universe,
and the metric of spacetime is divided into two parts:
\be   \label{metric}
g_{\alpha\beta}=g^{(0)}_{\alpha\beta}+h_{\alpha\beta},
\ee
where $g^{(0)}_{\alpha\beta}$ is the background spacetime metric,
$h_{\alpha\beta}$ is GW as small perturbations,
\be
\mid
h_{\alpha\beta}\mid\ll  1 .
\ee
As it stands,
the general definition of $t^{\alpha\beta}$ in Eq.(\ref{Landau-Lifshitz}) is
rather complex,
and, in practice, it can be simplified  for GW
with wavelength being smaller than the scale of curvature
of background spacetime.
In short wavelength limit
one can apply Brill-Hartle average \cite{BH,MTW,Isaacson1}.

There are three scales involved:
the wavelength  $\lambda$ of perturbations,
the scale $R$ of the curvature of background spacetime,
and a scale $L$ on which one takes average.
The condition for the short wavelength limit is
$\lambda\ll L \ll  R$.
The background metric $g^{(0)}_{\alpha\beta}$ is assumed
 to be a slowly varying function of spacetime,
whereas the perturbation $h_{\alpha\beta}$
changes significantly over small scales of space and time.
Let $\mathcal{A}$ be the characteristic amplitude
of $h_{\alpha\beta}$, $\mathcal{A}\ll 1$.
In vacuum, the energy density of perturbations is
 $\sim \frac{c^4}{G} (\frac{\mathcal{A}}{\lambda})^2$,
the curvature of background spacetime is  $\sim R^{-2}$.
 If the curvature of background spacetime
is induced by the energy density of perturbations, then, according to Einstein equation,
one has
$R^{-2}\sim (\frac{G}{c^4})(\frac{c^4}{G}) (\frac{\mathcal{A}}{\lambda})^2$,
i.e,
$\mathcal{A}\sim \frac{\lambda}{R}$.
By estimation of magnitude, one has
\[
g^{(0)}_{\mu\nu}\sim 1,\,\,\,\,g^{(0)}_{\mu\nu,\alpha}\sim \frac{1}{R},
\,\,\,\,g^{(0)}_{\mu\nu,\alpha\beta}\sim \frac{1}{R^2},
\]
\[
h_{\mu\nu}\sim \mathcal{A},\,\,\,\,
h_{\mu\nu,\alpha}\sim \mathcal{A}/\lambda\sim \frac{1}{R}.
\]
By Eq.(\ref{Landau-Lifshitz}),
 $t^{\alpha\beta}$ is quadratic in the first derivatives
of the metric.
The second order terms of perturbation in $t^{\alpha\beta}$,
denoted as $t^{(2)\alpha\beta}$,
should contain such terms of the following form
\[
hh,\,\,\, h\partial h,\,\,\, \partial h  \partial h,
\]
among them,
the terms like $\partial h  \partial h$ are dominant,
by estimation of the order of magnitude.
So we will only keep the terms like $\partial h \partial h$
in calculation.
Then Landau-Lifshitz' $t^{\alpha\beta}$ in Eq.(\ref{Landau-Lifshitz})
is reduced to
\ba \label{Landau-Lifshitz2}
16 \pi G\, t^{\alpha\beta} &=&A^{\alpha\beta}\,_\lambda A^{\lambda\mu}\,_\mu-
A^{\alpha\lambda}\,_{\lambda}A^{\beta\mu}\,_{\mu}
+\frac{1}{2}g^{\alpha\beta}g_{\lambda\mu}A^{\lambda\nu}\,_{\rho}A^{\rho\mu}\,_{\nu}\nonumber \\
&&-(g^{\alpha\lambda}g_{\mu\nu}A^{\beta\nu}\,_{\rho}A^{\rho\mu}\,_{\lambda}
+ g^{\beta\lambda}g_{\mu\nu}A^{\alpha\nu}\,_{\rho}A^{\rho\mu}\,_{\lambda})
+g^{\nu\rho}g_{\lambda\mu}A^{\alpha\lambda}\,_{\nu}A^{\beta\mu}\,_{\rho}
\nonumber \\
&&+\frac{1}{8}(2g^{\alpha\lambda}g^{\beta\mu}-g^{\alpha\beta}g^{\lambda\mu})
(2g_{\nu\rho}g_{\sigma\tau}-g_{\rho\sigma}g_{\nu\tau})
A^{\nu\tau}\,_{\lambda}A^{\rho\sigma}\,_{\mu}
\ea
where
\be
A^{\mu\nu}\,_\alpha \equiv (-g)^{-1/2}{\bf g}^{\mu\nu}\,_{,\alpha}
=g^{\mu\nu}\,_{,\alpha}+\frac{1}{2}g^{\mu\nu}g^{\rho\sigma}g_{\rho\sigma,\alpha}
\ee
contains the derivatives.
The terms of $t^{\alpha\beta}$  in Eq.(\ref{Landau-Lifshitz2})
are of the following form
\[
AA, \,    \, g g A A, \, \, g g g g A A.
\]
We expand $A^{\mu\nu}\,_\alpha$ to second order of perturbation:
\be \label{A}
A^{\mu\nu}\,_\alpha=A^{(0)\mu\nu}\,_\alpha +A^{(1)\mu\nu}\,_\alpha +A^{(2)\mu\nu}\,_\alpha,
\ee
where
\be
A^{(0)\mu\nu}\,_\alpha=g^{(0)\mu\nu}
\,_{,\alpha}+\frac{1}{2}g^{(0)\mu\nu}g^{(0)\rho\sigma}g^{(0)}\,_{\rho\sigma,\alpha},
\ee
\be \label{A^1}
A^{(1)\mu\nu}\,_\alpha
=-h^{\mu\nu}\,_{,\alpha}+\frac{1}{2}g^{(0)\mu\nu}g^{(0)\rho\sigma}h_{\rho\sigma,\alpha}
-\frac{1}{2}g^{(0)\rho\sigma}g^{(0)}_{\rho\sigma,\alpha}h^{\mu\nu}
-\frac{1}{2}g^{(0)\mu\nu}g^{(0)}_{\rho\sigma,\alpha}h^{\rho\sigma}
,
\ee
\ba
A^{(2)\mu\nu}\,_\alpha&=& \frac{1}{2}[g^{(0)\rho\sigma}g^{(0)}_{\rho\sigma,\alpha}h^{\mu}_{\gamma}h^{\gamma\nu}
+g^{(0)\mu\nu}g^{(0)}_{\rho\sigma,\alpha}h^{\rho}_{\gamma}h^{\gamma\sigma}
+g^{(0)}_{\rho\sigma,\alpha}h^{\rho\sigma}h^{\mu\nu}\nonumber\\
&&-g^{(0)\rho\sigma}h^{\mu\nu}h_{\rho\sigma,\alpha}
-g^{(0)\mu\nu}h^{\rho\sigma}h_{\rho\sigma,\alpha}+2h^{\mu}_{\gamma}h^{\gamma\nu}].
\ea
In short wavelength limit,
we keep only the second order of perturbation,  $\partial h  \partial h$,
in Eq.(\ref{Landau-Lifshitz2}).
Only the first two terms of $A^{(1)}$ contain $ \partial h$,
which are denoted by
\be \label{A1}
a^{(1)\mu\nu}\,_\alpha
\equiv -h^{\mu\nu}\,_{,\alpha}+\frac{1}{2}g^{(0)\mu\nu}g^{(0)\rho\sigma}h_{\rho\sigma,\alpha}.
\ee
More specifically,
$\partial h  \partial h$ actually come from $a^{(1)}a^{(1)}$.
So we have
\ba \label{Landau-Lifshitz3}
16 \pi G\, t^{(2)\alpha\beta}
&=&a^{(1)\alpha\beta}\,_\lambda a^{(1)\lambda\mu}\,_\mu-
a^{(1)\alpha\lambda}\,_{\lambda} a^{(1)\beta\mu}\,_{\mu}
  +\frac{1}{2}g^{(0)\alpha\beta}g^{(0)}_{\lambda\mu} a^{(1)\lambda\nu}\,_{\rho}a^{(1)\rho\mu}\,_{\nu}\nonumber\\
&& -[g^{(0)\alpha\lambda}g^{(0)}_{\mu\nu}a^{(1)\beta\nu}\,_{\rho} a^{(1)\rho\mu}\,_{\lambda}
+ g^{(0)\beta\lambda}g^{(0)}_{\mu\nu}a^{(1)\alpha\nu}\,_{\rho} a^{(1)\rho\mu}\,_{\lambda}]\nonumber\\
&&+g^{(0)\nu\rho}g^{(0)}_{\lambda\mu}a^{(1)\alpha\lambda}\,_{\nu} a^{(1)\beta\mu}\,_{\rho}
\nonumber \\
&&+\frac{1}{8}(2g^{(0)\alpha\lambda}g^{(0)\beta\mu}-g^{(0)\alpha\beta}g^{(0)\lambda\mu})\nonumber\\
&& \times (2g^{(0)}_{\nu\rho}g^{(0)}_{\sigma\tau}-g^{(0)}_{\rho\sigma}g^{(0)}_{\nu\tau})
a^{(1)\nu\tau}\,_{\lambda} a^{(1)\rho\sigma}\,_{\mu}.
\ea

Since the difference between  the ordinary derivatives
$h^{\mu\nu}\,_{,\alpha}$ and
covariant derivatives with respect to background metric $h^{\mu\nu}\,_{|\alpha}$
involves no terms like $\partial h$
and will not contribute to the expected terms like $\partial h\partial h $,
we replace the ordinary derivatives in Eq.(\ref{Landau-Lifshitz3})
by covariant derivatives with respect to background
metric $g^{(0)}_{\alpha\beta}$,
\ba \label{A2}
a ^{(1)\mu\nu}\,_\alpha
&=&-h^{\mu\nu}\,_{|\alpha}
+\frac{1}{2}g^{(0)\mu\nu}g^{(0)\rho\sigma}h_{\rho\sigma|\alpha}\nonumber\\
&=& -h^{\mu\nu}\,_{|\alpha}+\frac{1}{2}g^{(0)\mu\nu}h_{|\alpha},
\ea
where $h\equiv g^{(0)\mu\nu}h_{\mu\nu}$,
and the subscript  ``$_|$"  denotes
the covariant derivative using the background metric $g^{(0)}_{\alpha\beta}$,
and the indices are raised and lowered with the background metric.
In the gauge:
$h^{\alpha\beta}\, _{| \beta}=0$, $h=h^\alpha\, _\alpha=0$,
Eq(\ref{A2}) reduces to
\be
a^{(1)\mu\nu}\,_\alpha =-h^{\mu\nu}\,_{|\alpha},
\ee
and Eq.(\ref{Landau-Lifshitz3}) becomes
\ba \label{Landau-Lifshitz4}
16 \pi G\, t^{(2)\alpha\beta} &=& \frac{1}{2}g^{(0)\alpha\beta}g^{(0)}_{\lambda\mu}h^{\lambda\nu}\,_{|\rho}h^{\rho\mu}\,_{|\nu}
-h^{\rho\mu|\alpha}h^\beta_{\mu|\rho}-h^{\rho\mu|\beta}h^\alpha_{\mu|\rho}\nonumber\\
&&+h^{\beta\mu|\nu}h^\alpha_{\mu|\nu}+\frac{1}{2}h^{\rho\sigma|\alpha}h_{\rho\sigma}\,^{|\beta}
-\frac{1}{4}g^{(0)\alpha\beta}h^{\rho\sigma|\mu}h_{\rho\sigma|\mu}.
\ea
This expression is still clumsy,
and can be simplified in many applications
using an averaging process adopted in Refs. \cite{ADM,BH,Isaacson1}.
Whenever  regions of scale $L$ are large enough to contain
many wavelengths,
it is natural and necessary to employ the average.
{The idea is similar  to finding electric fields in
a macroscopic dielectric.
One is not interested in fine details of
stochastic electric fields fluctuating over molecular scales,
but only need to know electric fields
averaged over a region large enough to contain many
molecules and yet small compared with the size of the dielectric.
In general,
GW is similar to this.
$h_{\mu\nu}$ consists of all sorts of modes
of various wavelengths $\lambda$,
and those with  $\lambda \ll L$ are intractable,
practically regarded as being stochastic.
We will take average of
 Eq.(\ref{Landau-Lifshitz4}) over a spacetime region of scale $L$.

We only sketch the main idea of this kind of averaging procedure,
omitting the technical details concerning the tensorial nature,
which can be found in Ref.\cite{Isaacson1}.
In right hand side of Eq.(\ref{Landau-Lifshitz4})
there are six terms of the form $\partial h \partial h$.
Consider a gradient $H _{|\sigma}$
composed of quadratic of  $\partial h \partial h$.
The average of $H\, _{|\sigma}$  over the region
can be schematically defined as
\[
 < H_{|\sigma}(x)> =
  \int  H _{|\sigma}(x')f(x,x')d^4x',
\]
where $f(x,x')$ is a weighting (window) function which
falls smoothly to zero when $|x-x'|>L$,
and satisfies a normalization $\int f(x,x')d^4 x=1$.
Then,
\[
 < H _{|\sigma}(x)> =
  \int [\, ( H f )_{|\sigma} -H  f_{|\sigma}  \, ] d^4x'.
\]
The first term is actually a surface integral $\int_{S} H f d^3x'$
which can be dropped as $f|_{S}=0$ on the boundary surface,
and the second term contains $f_{|\sigma}  \sim f/L $.
Note that for GW,
a derivative
$h^{\mu\nu} \, _{|\sigma}\sim h^{\mu\nu}/\lambda $,
so $H _{|\sigma}\sim  H /\lambda$.
Thus,
the overall effect of averaging on the gradient $H_{|\sigma}$
is such that
$H /\lambda \rightarrow H /L $.
Since we assume $\lambda /L \ll 1$,
the  divergence in average is vanishing $ < H_{|\sigma}(x)> \simeq 0$.
From this result follow  the rules of
Brill-Hartle average  (see \S35.15 in Ref.\cite{MTW}):
\begin{itemize}
\item   Gradients average out to zero; e.g.,
$<(h^{\gamma\delta}\,_{|\alpha}h_{\mu\nu})_{|\beta}>=0$;
\item   One can freely integrate by parts, flipping derivatives from
one $h$ to the other; e.g.,
$<h^{\gamma\delta} h_{\mu\nu|\alpha\beta}>
=<-h^{\gamma\delta}\,_{|\beta} h_{\mu\nu|\alpha}>$;
\item   Covariant derivatives commute; e.g.,
   $<h^{\gamma\delta} h_{\mu\nu|\alpha\beta}>
          =<h^{\gamma\delta} h_{\mu\nu|\beta\alpha}>$.
\end{itemize}
The last rule is derived with the help
of the equation of motion for $h_{\mu\nu}$  \cite{Isaacson1}
\be
h_{\mu\nu|\rho}\,^\rho+2R^{(0)}_{\sigma\nu\mu\rho}h^{\rho\sigma}
+R^{(0)}_{\sigma\mu}h^\sigma_\nu
+R^{(0)}_{\sigma\nu}h^\sigma_\mu=0.
\ee
Applying these rule to Eq.(\ref{Landau-Lifshitz4}),
only one term $(1/2)h^{\rho\sigma|\alpha}h_{\rho\sigma}\,^{|\beta}$
remains,
}
and
we arrive at the energy momentum tensor for gravitational wave
\be \label{t2}
 <t^{(2)\alpha\beta}>=t^{BH}\, ^{\alpha\beta}
  \equiv \frac{1}{32 \pi G}<h^{\rho\sigma|\alpha}h_{\rho\sigma}\,^{|\beta}>,
\ee
This expression has a desired property that
it is a gauge invariant \cite{Isaacson1,MTW}
and a physical observable.
It has been commonly  used in studies of gravitational radiation.
For RGW in an expanding universe,
Eq.(\ref{t2}) can be used,
as long as the  wavelengths under consideration
are much shorter than the Hubble radius $1/H_0\sim 3000$Mpc.
Otherwise, one need use Eq.(\ref{Landau-Lifshitz4}).

\begin{center}
{\bf 4. The Effective $t^{\alpha\beta}_{eff}$}
\end{center}

Another definition of the energy-momentum pseudo-tensor
of GW has been used in literature,
called the effective EMPT \cite{Isaacson1,MTW,Mukhanov1,Mukhanov2,Giovannini2}.
We now examine its relevant properties.
Consider a curved spacetime in Eq.(\ref{metric})
filled with  matter.
Einstein equation
\be
G^{\alpha\beta}=8 \pi G T^{\alpha\beta}
\ee
is expanded to the second order of perturbation,
\be
G^{(0)\alpha\beta}+G^{(1)\alpha\beta}+G^{(2)\alpha\beta}=8\pi G
[T^{(0)\alpha\beta}+T^{(1)\alpha\beta}+T^{(2)\alpha\beta}].
\ee
After rearranging some terms in the above equation we have
\be \label{G}
G^{(0)\alpha\beta}=8\pi G
[T^{(0)\alpha\beta}+T^{(2)\alpha\beta}-\frac{1}{8\pi
G}G^{(2)\alpha\beta}]+[G^{(1)\alpha\beta}-8\pi GT^{(1)\alpha\beta}].
\ee
The first order part of perturbation satisfies
their field equation
\be
G^{(1)\alpha\beta} =8\pi GT^{(1)\alpha\beta},
\ee
which will be the main subject of cosmological perturbations.
Then  Eq.(\ref{G}) reduces to
\be \label{E'equation}
G^{(0)\alpha\beta}=
8\pi G [T^{(0)\alpha\beta}+ T^{(2)\alpha\beta}+t^{\alpha\beta}_{eff}],
\ee
where
\be
t^{\alpha\beta}_{eff}=-\frac{1}{8\pi  G} G^{(2)\alpha\beta}
\ee
is the effective EMPT \cite{Isaacson1,Mukhanov1,Mukhanov2}.
As the second order part,
 $G^{(2)\alpha\beta}$  consists  of quadratic terms of metric perturbations
 and will affect background spacetime,
known as the ``backreaction" effect
\cite{MTW,Mukhanov1,Mukhanov2,Isaacson1}.
Note that
the second order matter term $T^{(2)\alpha\beta}$
also appears and affects the background spacetime as well.
In practice, it is small and can be neglected.
Explicitly,
\ba   \label{teff}
-8\pi  G t^{\alpha\beta}_{eff}
&=&R^{(2)\alpha\beta}-\frac{1}{2}g^{(0)\alpha\beta}g^{(0)}_{\mu\nu}R^{(2)\mu\nu}
+\frac{1}{2}[g^{(0)}_{\mu\nu}h^{\alpha\beta}-g^{(0)\alpha\beta}h_{\mu\nu}]R^{(1)\mu\nu} \nonumber\\
&&+\frac{1}{2}[R^{(0)\mu\nu}h_{\mu\nu}h^{\alpha\beta}
-R^{(0)}h^{\alpha\rho}h^{\beta}_{\rho}],
\ea
where the perturbed Ricci tensors are
\be
R^{(1)\mu\nu}=g^{(0)\mu\alpha}g^{(0)\nu\beta}R^{(1)}_{\alpha\beta}
-[g^{(0)\mu\alpha}h^{\nu\beta}+g^{(0)\nu\beta}h^{\mu\alpha}]R^{(0)}_{\alpha\beta},
\ee
\ba
R^{(2)\mu\nu}&=&g^{(0)\mu\alpha}g^{(0)\nu\beta}R^{(2)}_{\alpha\beta}
-[g^{(0)\mu\alpha}h^{\nu\beta}+g^{(0)\nu\beta}h^{\mu\alpha}]R^{(1)}_{\alpha\beta}\nonumber\\
&&+[R^{(0)\mu\alpha}h^{\nu\beta}+R^{(0)\nu\alpha}h^{\mu\beta}]h_{\alpha\beta}
+R^{(0)}_{\alpha\beta}h^{\mu\alpha}h^{\nu\beta}
\ea
with (see Eq.(35.58) in Ref.\cite{MTW})
\be
R^{(1)}_{\mu\nu}=\frac{1}{2}
(-h_{|\mu\nu}-h_{\mu\nu|\alpha}\,^\alpha
+h_{\alpha\mu|\nu}\,^\alpha+h_{\alpha\nu|\mu}\,^\alpha),
\ee
\ba
R^{(2)}_{\mu\nu}&=&\frac{1}{2}[\frac{1}{2}h_{\alpha\beta|\mu}h^{\alpha\beta}\,_{|\nu}
+h^{\alpha\beta}(h_{\alpha\beta|\mu\nu}+h_{\mu\nu|\alpha\beta}-h_{\alpha\mu|\nu\beta}-h_{\alpha\nu|\mu\beta})\nonumber\\
&&+h_{\nu}\,^{\alpha|\beta}(h_{\mu\alpha|\beta}-h_{\mu\beta|\alpha})\nonumber\\
&&-(h^{\alpha\beta}\,_{|\beta}
-\frac{1}{2}h^{|\alpha})(h_{\alpha\mu|\nu}+h_{\alpha\nu|\mu}-h_{\mu\nu|\alpha})].
\ea
Note that  $t^{\alpha\beta}_{eff}$
is not covariantly conserved, $t_{eff}^{\alpha\beta}\, _{|\beta} \neq0$,
using the background metric.
As it stands in Eq.(\ref{teff}),
$t^{\alpha\beta}_{eff}$ is
different from the expression in Eq.(\ref{Landau-Lifshitz4})
by keeping only  quadratic $\partial h \partial h $ part
of Landau-Lifshitz' $ t^{\mu\nu}_{LL}$,
and is also clumsy for a general curved background spacetime.
Similar to what
we have done for Landau-Lifshitz' $t^{\alpha\beta}$
in the last section,
one takes the average by applying the Brill-Hartle rule to Eq.(\ref{teff}),
yielding \cite{Isaacson1}
\be \label{tBH}
<t^{\alpha\beta}_{eff}>= \frac{1}{32\pi
G}<h^{\rho\sigma|\alpha}h_{\rho\sigma}\,^{|\beta}>,
\ee
where the gauge condition
$h^{\alpha\beta}\, _{| \beta}=0$, $h^\alpha\, _\alpha=0$,
are used.
Thus, in the short wavelength limit after the Brill-Hartle average,
the effective $t^{\alpha\beta}_{eff}$
defined in Eq.(\ref{teff}) is equivalent to
Landau and Lifshitz'  $t^{\alpha\beta}$,
both leading to the same expression.
One sometimes extrapolates to use Eq.(\ref{teff})
in the model of bouncing universe
to investigate the backreaction of gravitons upon
 the background spacetime  \cite{Giovannini2}.

In the long wavelength limit \cite{Mukhanov1,Mukhanov2,Giovannini2},
the wavelength of perturbations is comparable to, or even larger than
the scale of curvature of background spacetime.
This can be relevant for study of the very early universe
when the wavelengths of interest are even bigger than the horizon.
Then the Brill-Hartle average procedure is not available.
One may try to do  spatial average of
the EMPT in a fixed time slice of spacetime.
However, there is a problem of gauge invariance with this method,
and the issue  is not settled down \cite{Mukhanov1,Unruh}.

\begin{center}
{\bf 5. Pressure of RGW }
\end{center}

We consider a spatial flat homogeneous and isotropic RW  spacetime
 filled with only RGW.
The metric is given by Eq.(\ref{metric})
with the background $g^{(0)}_{\alpha\beta}=a^2(\tau)diag (1,-1,-1,-1)$,
and the perturbations in the synchronous coordinate are given by
\be \label{hij}
h_{\alpha \beta} \equiv  \begin{array}{l}
\left( {\begin{array}{*{20}c}
   0 & 0  \\
   0 & -a^2(\tau)h_{ij} \\
\end{array}} \right), \,\,\,\, i,j=1,2,3.
\end{array}
\ee
In the following as RGW
 $h_{ij}$ stands for what appear in Eq.(\ref{hij}) and satisfies the same
TT condition $h^i_i=\partial_i h^i_j=0$ as in Eq.(\ref{TT}).
For this spacetime,
a direct calculation of $t^{\alpha\beta}_{eff}$ given by Eq.(\ref{teff})
yields
the energy density and pressure for RGW
\be \label{rho}
\rho_{gw}=t^0_0= \frac{1}{8\pi G a^2}[\mathcal{H} h^{'}_{kl} h^{kl}
+\frac{1}{8}(\partial_m h_{kl} \partial^m h^{kl}
+h^{'}_{kl} h^{' kl})],
\ee
\be \label{p}
p_{gw}=-\frac{1}{3}t^{i}_{i}
= \frac{1}{24\pi G a^2}[\frac{7}{8}\partial_m h_{kl} \partial^m h^{kl}
-\frac{5}{8} h^{'}_{kl} h^{' kl}],
\ee
where $\mathcal{H}=a'(\tau)/a(\tau)$ and
the prime denotes the time derivative with respect to the conformal time.
We drop the subscript `` eff " for simplicity.
In deriving the expressions of
$\rho_{gw}$ and $p_{gw}$,
a total derivative has been dropped, respectively.
No average is taken on the expressions (\ref{rho}) and (\ref{p}),
which is different from  those given by the averaged expression
in Eq.(\ref{tBH}).
As mentioned in last section,
the effective $t^{\alpha\beta}$ under discussion is not covariantly conserved.
To get a conservation equation of RGW,
Ref.\cite{Giovannini2}
expands the Bianchi identity $G^\mu\,_{\nu\, ;\mu}=0$
to the the second order of perturbation:
\be
0= G^\mu\,_{\nu\, ;\mu}=(G^\mu\,_{\nu\, ;\mu})^{(0)}
+(G^\mu\,_{\nu\, ;\mu})^{(1)}+(G^\mu\,_{\nu\, ;\mu})^{(2)},
\ee
requires each order should vanish respectively.
Then the second order part
\be
(G^\mu\,_{\nu\, ;\mu})^{(2)}=0,
\ee
which reads as
\be  \label{ece}
\frac{\partial \rho_{gw}}{\partial \tau}
+3 \mathcal{H} (\rho_{gw}+\mathcal{P}_{gw})=0,
\ee
where the pressure \cite{Giovannini2},
\be  \label{Pressure}
\mathcal{P}_{gw} \equiv p_{gw}
    +\frac{\mathcal{H}^2-\mathcal{H}^{'}}{24\pi G \mathcal{H} a^2}
     h^{'}_{kl} h^{ kl}.
\ee

We will take a different approach to the conservation equation
of RGW,
and trace the origin of $\mathcal{P}_{gw} $.
Taking covariant derivative on both sides of Einstein equation (\ref{E'equation})
using the background metric $g^{(0)}_{\alpha\beta}$
leads to
\be
G^{(0)\mu}\, _{\nu | \mu}=8 \pi G [T^{(0)\mu}\,_\nu+t^\mu_\nu]\, _{| \mu}.
\ee
By the Bianchi identity $G^{(0)\mu}\, _{\nu | \mu}=0$,
we have
\be \label{concon}
[T^{(0)\mu}\,_\nu+t^\mu_\nu]\, _{| \mu}  =0,
\ee
which tells that the total energy and momentum tensor, i.e,
the sum of matter  and  GW,
is covariantly conserved, not $t^\mu_\nu$ itself.
Only in absence of matter with $T^{(0)\mu}\,_\nu=0$,
the EMPT of RGW is covariantly conserved with
\be \label{teq}
t^{\mu}\, _{\nu | \mu}=0,
\ee
which, for the  RW spacetime, explicitly reads as
\be \label{eqrho}
\frac{\partial \rho_{gw}}{\partial \tau}+3 \mathcal{H} (\rho_{gw}+p_{gw})=0,
\ee
which is commonly used in literature on RGW.
It is straightforward to show that
the covariant conservation of the Brill-Hartle averaged tensor,
\be
t^{BH}\, _{\nu}^{\mu}\,_{|\mu}=0
\ee
also leads to Eq.(\ref{eqrho}).
We would refer to Eq.(\ref{eqrho}) as the ``canonical'' equation of
energy conservation of RGW.
It is noticed that Eq.(\ref{ece}) differs from Eq.(\ref{eqrho})
with $\mathcal{P}_{gw}$ replacing $p_{gw}$.

In general, when the matter is present, $T^{(0)\mu}\,_\nu \ne 0$,
Eq.(\ref{concon}) gives
\be \label{tT}
t^{\mu}\, _{\nu | \mu} =-T^{(0)\mu}\, _{\nu | \mu},
\ee
where the matter term $-T^{(0)\mu}\, _{\nu | \mu}$
on the right hand side as an extra term
can be interpreted as the interaction between GW and matter.
We give  a detailed calculation of this term.
Starting with
the covariant conservation equation of energy momentum tensor for matter,
\be \label{ccem}
T^{(0)\mu}\, _{\nu ; \mu}=0,
\ee
expanding it to second order of perturbations,
and taking spatial average,
one has
\be
<T^{(0)\mu}\, _{\nu ; \mu}>=T^{(0)\mu}\, _{\nu | \mu}
+<\Gamma^{(2)\mu}\,_{\alpha\mu}>T^{(0)\alpha}\,_{\nu}
-<\Gamma^{(2)\alpha}\,_{\nu\mu}>T^{(0)\mu}\,_{\alpha}=0,
\ee
so that
\be \label{emccem}
T^{(0)\mu}\, _{\nu | \mu}=-<\Gamma^{(2)\mu}\,_{\alpha\mu}>T^{(0)\alpha}\,_{\nu}
+<\Gamma^{(2)\alpha}\,_{\nu\mu}>T^{(0)\mu}\,_{\alpha}.
\ee
This result tells that $T^{(0)\mu}\, _{\nu | \mu}$ represents
a coupling between the metric perturbation
and the background matter as given by
the expression on the right hand side.
Since the difference between $T^{(0)\alpha}\,_{\nu}$
and $G^{(0)\alpha}\,_{\nu}/8\pi G $ is of second order,
$T^{(0)\alpha}\,_{\nu}$ on the right hand side of Eq.(\ref{emccem})
can be replaced by $G^{(0)\alpha}\,_{\nu}$.
For the $\nu=0$ component, this leads to
\ba \label{ecceg}
T^{(0)\mu}\, _{0 | \mu}
&=& \frac{1}{8\pi G } (-<\Gamma^{(2)\mu}\,_{\alpha\mu}>G^{(0)\alpha}\,_{0}
+<\Gamma^{(2)\alpha}\,_{0\mu}>G^{(0)\mu}\,_{\alpha}).
\ea
By calculation,
\[
G^0_{0}=3(\frac{a'}{a^2})^2,
\]
\[
G^i_j= (2\frac{a''}{a^3}-(\frac{a'}{a^2})^2)\delta^i_j,
\]
\[
<\Gamma^{(2)i}\,_{j0}>=-\frac{1}{2}h^{ik}h'_{kj},
\]
\[
<\Gamma^{(2)k}\,_{ij}>= \frac{1}{2}h^{il}
(h^{jk},_l -h_{jl},_k- h_{kl},_j),
\]
one obtains
\be \label{t0}
T^{(0)\mu}\, _{0 | \mu}
     =\frac{\mathcal{H}^2-\mathcal{H}^{'}}{8\pi G  a^2}
                   h^{'}_{kl} h^{ kl}.
\ee
In this form , $T^{(0)\mu}\, _{\nu | \mu}$ is
a coupling between the gravitational waves
and the background  spacetime  of expanding universe.
Thus, Eq.(\ref{tT}) is cast  into the form
\be \label{eqrho2}
\frac{\partial \rho_{gw}}{\partial \tau}+3 \mathcal{H} (\rho_{gw}+p_{gw})
     =-\frac{\mathcal{H}^2-\mathcal{H}^{'}}{8\pi G  a^2}
                   h^{'}_{kl} h^{ kl},
\ee
which is the same as Eq.(\ref{ece}) with
\be
\mathcal{P}_{gw} =p_{gw} +\frac{1}{3\mathcal{H} }T^{(0)\mu}\, _{0| \mu}.
\ee

By now,
the origin of the pressure $\mathcal{P}_{gw}$ is clear.
In presence of matter,
the equation of energy conservation
can still be written in a form of continuity equation as Eq.(\ref{ece}),
where, for the pressure,
one has to use $\mathcal{P}_{gw}$
to take into account the interaction between GW and matter.
Since the term
$\frac{\mathcal{H}^2-\mathcal{H}^{'}}{24\pi G \mathcal{H} a^2}
     h^{'}_{kl} h^{ kl}$
as  the difference $\mathcal{P}_{gw}- p_{gw}$
is a typical viscosity term for RGW since it
 involves the time derivative  $h'_{kl}$.
The prefactor $\frac{\mathcal{H}^2-\mathcal{H}^{'}}{24\pi G  a^2}$
determines the direction and rate of energy interchange
between RGW and matter.
But, for RGW whose wavelengths are within the horizon of the expanding universe,
this term is a small correction
to $p_{gw}\sim \frac{1}{96\pi G a^2} h^{'}_{kl} h^{' kl}$,
since the rates $\mathcal{H}$, $\mathcal{H}'/\mathcal{H}$
are smaller than the frequencies of RGW of interest,
and therefore
$\frac{\mathcal{H}^2-\mathcal{H}^{'}}{\mathcal{H} } |h^{ kl}| \ll | h^{' kl}|$.
In short-wavelength limit,
the correction term is recognized as being
one order of $O(\lambda/R)$  higher than $p_{gw}$.

Moreover,
for the inflationary stage with the scale factor $a(\tau)\propto 1/\tau$
\be
\mathcal{H}^2-\mathcal{H}^{'}=0,
\ee
one has $\mathcal{P}_{gw}=p_{gw}$.
For a general expansion stage with $a(\tau)\propto \tau^{\alpha}$,
one has
\be
\frac{\mathcal{H}^2-\mathcal{H}^{'}}{\mathcal{H}}
       = \frac{\alpha+1}{\tau} ,
\ee
which decreases quickly with the expansion,
and ${\mathcal{P}_{gw}}$ effectively approaches to $p_{gw}$.
This is true for the radiation-dominated stage ($\alpha =1$)
and the matter-dominated stage ($\alpha = 2$).
Putting all these consideration together,
for whole history of cosmic expansion since inflation,
the difference  $\mathcal{P}_{gw}-p_{gw}$ is negligibly small,
and one can amply use Eq.(\ref{eqrho}) for energy conservation
in practical studies of RGW.
Only in the pre-inflationary stage
the difference  $\mathcal{P}_{gw}-p_{gw}$
can be significant for RGW of long wavelengths  \cite{Giovannini2}.

\begin{center}
{\bf 6. Effective EMPT and  Wave Equation }
\end{center}

For the effective EMPT,
we now explicitly show that the energy conservation
is equivalent the field equation of RGW in the expanding RW spacetime.
The field equation for RGW is
\be  \label{gwe}
h^{''}_{kl}+2 \mathcal{H} h^{'}_{kl}-\partial^m \partial_m h_{kl}=0,
\ee
which is commonly used in literature.
The equation of energy conservation is given in Eq.(\ref{ece}).
Taking time derivative of $\rho_{gw}$  as given in Eq.(\ref{rho}),
we have
\ba
\frac{\partial \rho_{gw}}{\partial \tau}
&=&-\frac{2 a'}{8\pi G a^3}[\mathcal{H} h^{'}_{kl} h_{kl}
+\frac{1}{8}(\partial_m h_{kl} \partial^m h^{kl}
+h^{'}_{kl} h^{' kl})]  \nonumber\\
&&+\frac{1}{8\pi G a^2}[\mathcal{H}^{'} h^{'}_{kl} h_{kl}
+\mathcal{H} h^{''}_{kl} h_{kl}+\mathcal{H} h^{'}_{kl} h^{'}_{kl}+\frac{1}{4}(\partial_m h^{'}_{kl} \partial^m h^{kl}+h^{''}_{kl} h^{' kl})]\nonumber\\
&=&-2\mathcal{H}\rho_{gw}+\frac{1}{8\pi G a^2}[\mathcal{H}^{'} h^{'}_{kl} h_{kl}+\mathcal{H} h^{''}_{kl} h_{kl}+\mathcal{H} h^{'}_{kl} h^{'}_{kl}\nonumber\\
&&+\frac{1}{4}(\partial_m h^{'}_{kl} \partial^m h^{kl}+h^{''}_{kl} h^{' kl})].\nonumber\\
\ea
Using  Eq.(\ref{rho}) for $\rho_{gw}$
and Eq.(\ref{Pressure}) for $\mathcal{P}_{gw}$
to calculate $\mathcal{H} \rho_{gw}+3 \mathcal{H}\mathcal{P}_{gw}$ yields
\ba
\mathcal{H} \rho_{gw} +3 \mathcal{H}\mathcal{P}_{gw}
&=&\mathcal{H} \rho_{gw}+3 \mathcal{H} p_{gw}+\frac{\mathcal{H}^2-\mathcal{H}^{'}}{8\pi Ga^2} h^{'}_{kl} h^{ kl}\nonumber\\
&=&\frac{1}{8\pi G a^2}[(2\mathcal{H}^2-\mathcal{H}^{'})h^{'}_{kl} h_{kl}+\mathcal{H}\partial_m h_{kl} \partial^m h^{kl}\nonumber\\
&&-\frac{1}{2} \mathcal{H} h^{'}_{kl} h^{' kl}].
\ea
Adding these together,  by using
$\partial_m h^{'}_{kl} \partial^m h^{kl}=-  h^{'}_{kl}\partial_m \partial^m h^{kl}$
 plus a surface term which is dropped,
we finally arrive at
the following result:
\be \label{ECE}
\frac{\partial \rho_{gw}}{\partial \tau}
           +3 \mathcal{H} (\rho_{gw}+\mathcal{P}_{gw})
  =\frac{1}{8\pi G a^2}(\frac{1}{4}h^{' kl}+\mathcal{H} h^{kl})
        (h^{''}_{kl}+2 \mathcal{H} h^{'}_{kl}-\partial^m \partial_m h_{kl}).
\ee
From this equality, one sees that
the wave equation (\ref{gwe})
 implies the  equation of energy conservation (\ref{ece}),
and vice versus.
Considering that for most of stages of cosmic expansion
$\mathcal{P}_{gw}\simeq p_{gw}$,
one can amply use either the wave equation (\ref{gwe}) or
the conservation equation (\ref{eqrho})
to study RGW.

In the special case of  $\mathcal{H} =0$,
Eq.(\ref{ECE} ) reduces to (choosing $a=1$)
\be
\frac{\partial \rho_{gw}}{\partial \tau}
  =\frac{1}{32\pi G }h^{' kl} (h^{''}_{kl}-\partial^m \partial_m h_{kl}),
\ee
the same as Eq.(\ref{con}) for Landau-Lifshitz' case.

\begin{center}
{\bf 7. Conclusion}
\end{center}

The wave equation of GW is shown to be equivalent to
the equation of energy conservation for Landau-lifshitz' definition of
the pseudo-tensor $t^{\alpha\beta}$ in flat spacetime.
It is explicitly demonstrated that,
in the short wavelength limit after Brill-Hartle average,
both Landau-lifshitz'  $t^{\alpha\beta}$ and
the effective  $t^{\alpha\beta}_{eff}$
lead to the gauge-invariant, concise expression
$t^{BH}_{\alpha\beta}=\frac{1}{32\pi G}<h^{\mu\nu}\,_{|\alpha}h_{\mu\nu|\beta}>$,
which is often used in study of gravitational waves.

For RGW in the RW spacetime,
the two forms of pressures  $p_{gw}$ and $\mathcal{P}_{gw}$
are analyzed,
and their difference
is traced to be $\mathcal{P}_{gw}-p_{gw}= \frac{1}{3\mathcal{H} }T^{(0)\mu}\, _{0| \mu}$.
When matter is present,
the energy conservation of RGW
can still be written as a form of continuity equation
$\frac{\partial \rho_{gw}}{\partial \tau}+3{ \mathcal{H}} (\rho_{gw}+\mathcal{P}_{gw})=0$,
where the pressure $\mathcal{P}_{gw}$
takes into account the interaction between RGW and matter.
For  wavelengths within the horizon,
the difference $\mathcal{P}_{gw}-p_{gw}$ is small,
compared to $p_{gw}$,
and decreases with  the cosmic expansion.
During the de Sitter stage $\mathcal{P}_{gw}-p_{gw}=0$.

We have also explicitly demonstrated that the wave equation
$h^{''}_{kl}+2 \mathcal{H} h^{'}_{kl}-\partial^m \partial_m h_{kl}=0$
is equivalent to
$\frac{\partial \rho_{gw}}{\partial \tau}+3 \mathcal{H} (\rho_{gw}+\mathcal{P}_{gw})=0$.
Nevertheless,
since $\mathcal{P}_{gw}  \simeq  p_{gw}$ after inflation,
one can amply use either the wave equation
or the ``canonical'' equation
$\frac{\partial \rho_{gw}}{\partial \tau}+3 \mathcal{H} (\rho_{gw}+p_{gw})=0$
in study of RGW for practical purposes.

\

ACKNOWLEDGMENT:
 Y. Zhang's research work has been supported by
the CNSF No.11073018, SRFDP, and CAS.

\end{document}